\documentclass{article}
\usepackage[utf8]{inputenc}
\usepackage[normalem]{ulem}
\usepackage{graphicx} 
\usepackage{authblk}
\usepackage{xcolor}
\usepackage{tablefootnote}
\usepackage{cite}
\usepackage{amsmath,amssymb,amsfonts}
\usepackage{mathtools}
\usepackage{algorithmic}
\usepackage{textcomp}
\usepackage{relsize}

\usepackage[font=small,labelfont=bf]{caption}
\usepackage[square,numbers]{natbib}
\usepackage{hyperref}

\usepackage{graphicx}
\usepackage{dcolumn}
\usepackage{bm}

\usepackage[utf8]{inputenc}
\usepackage[T1]{fontenc}
\usepackage{mathptmx}
\usepackage{etoolbox}
\usepackage{xcolor}
\usepackage[table]{xcolor}
\usepackage{multirow}
\usepackage{booktabs}
\usepackage{mathrsfs}
\usepackage[mathcal]{euscript}
\usepackage{cite}
\usepackage{makecell}
\usepackage{siunitx}

\newcounter{bla}

\newcommand{\be}{\begin{equation}}

\newcommand{\ee}{\end{equation}}
\definecolor{codegreen}{rgb}{0,0.6,0}
\definecolor{codegray}{rgb}{0.5,0.5,0.5}
\definecolor{codepurple}{rgb}{0.58,0,0.82}
\definecolor{backcolour}{rgb}{0.95,0.95,0.92}

\usepackage{listings}
\definecolor{mGreen}{rgb}{0,0.6,0}
\definecolor{mGray}{rgb}{0.5,0.5,0.5}
\definecolor{mPurple}{rgb}{0.58,0,0.82}
\definecolor{backgroundColour}{rgb}{0.95,0.95,0.92}
\lstdefinestyle{CStyle}{
    backgroundcolor=\color{backgroundColour},   
    commentstyle=\color{mGreen},
    keywordstyle=\color{magenta},
    numberstyle=\tiny\color{mGray},
    stringstyle=\color{mPurple},
    basicstyle=\footnotesize,
    breakatwhitespace=false,         
    breaklines=true,                 
    captionpos=b,                    
    keepspaces=true,                 
    numbers=left,                    
    numbersep=5pt,                  
    showspaces=false,                
    showstringspaces=false,
    showtabs=false,                  
    tabsize=2,
    language=C
}

\title{Data-Driven Reconstruction of Spatially Resolved Electron and Ion Energy Distributions from Macroscopic Plasma Quantities with Deep Neural Networks}
\author{Libin Varghese$^1$, Kaushik Prajapati$^1$, Bhaskar Chaudhury$^1$}

\affil{$^1$Group in Computational Science and HPC, Dhirubhai Ambani University (formerly DA-IICT), Gandhinagar, INDIA. \texttt{bhaskar\_chaudhury@dau.ac.in}}

\date{ }
\providecommand{\keywords}[1]{\textbf{\textit{Keywords-}} #1}

\begin{document}
\maketitle

\begin{abstract}
Spatially resolved electron and ion energy distribution functions (EEDFs/IEDFs) provide essential kinetic information about low-temperature plasmas (LTPs) and play a central role in determining transport, chemical reaction rates, and plasma surface interactions.  While kinetic simulations directly resolve these distributions, experimental measurements remain challenging and are often invasive, spatially limited, or require assumptions regarding the distribution shape such as a Maxwellian, that may not always hold  under LTP conditions. However, several macroscopic plasma observables can be measured non-invasively using advanced diagnostic techniques, providing spatially resolved information about the plasma state. An important inverse problem is therefore whether readily measurable macroscopic plasma quantities contain sufficient information to reconstruct the underlying kinetic state. In this work, we investigate this problem by learning a nonlinear mapping from spatially resolved macroscopic plasma observables to the corresponding spatially resolved EEDFs/IEDFs using a deep learning framework. Paired datasets comprising 2D macroscopic observables and spatially resolved EDFs are generated using 2D-3V Particle-in-Cell Monte-Carlo-Collision (PIC-MCC) simulations. Three representative learning paradigms, a convolutional encoder-decoder (U-Net), a Fourier Neural Operator (FNO), and a graph-based MeshGraphNet, are employed in this study to learn this inverse mapping. The predicted EDFs reproduce both bulk plasma and sheath characteristics with good agreement to the PIC-MCC reference data, with the FNO providing the best overall performance. Beyond conventional metrics, physics-based validation demonstrates that the reconstructed EDFs accurately recover the corresponding density and temperature, and rate coefficients, confirming their physical consistency. These results demonstrate that macroscopic plasma observables encode sufficient information to infer important kinetic properties in LTPs, providing a potential foundation for surrogate kinetic modeling and next-generation plasma diagnostics.

\end{abstract}


\section{Introduction} \label{sec:intro}
Low temperature plasmas (LTPs) are partially ionized gases characterized by non-equilibrium conditions where energetic electrons play a dominant role in sustaining the discharge and driving ionization, excitation, and chemical reactions\cite{fridman,ltpfoundation}. A distinctive feature of LTPs is the large difference between electron and ion temperature, making LTPs attractive for diverse applications including semiconductor manufacturing, space exploration, fusion research, plasma assisted combustion, additive manufacturing and many more \cite{adamovich20222022,adamovich20172017}. Plasma characterization and diagnostic methods remain essential for optimizing LTP devices\cite{hippler2001low}. Plasma parameters are generally classified into macroscopic parameters (fluid observables) and microscopic kinetic quantities (such as energy distribution functions) \cite{demidov2002electric}. Energy distribution functions (EDFs) are central to plasma kinetics, chemistry, and reaction rates\cite{Kolobov2}.
Significant progress has been made in measuring plasma parameters, such as density, temperature, and EDFs, using both invasive and non-invasive diagnostic techniques \cite{adamovich20172017,eedf1,hanna2020investigating}. In addition to experimental diagnostics, considerable effort has been devoted to the theoretical and computational investigation of electron and ion energy distribution functions (EEDFs and IEDFs) through kinetic models\cite{sredf, hagelar}. Approaches based on the Boltzmann equation, Monte Carlo simulations, and Particle-In-Cell (PIC) methods have played a central role in advancing and understanding non-equilibrium plasma behavior, transport processes, and reaction kinetics\cite{alvesreview,donko,boeuf2025stratification}. Macroscopic parameters such as plasma density and average electron temperature are obtained from moments or integrals of the underlying microscopic EEDF \cite{kolobov2019electron}. 

The EEDF governs electron-impact reaction rates, transport coefficients, ionization, excitation, and other kinetic processes in LTPs \cite{Kolobov2}. Under the strongly non-equilibrium conditions, the EEDF often departs from the Maxwellian form owing to the combined effects of electron heating, transport, and collisional processes \cite{kolobov2019electron}. Similarly, the IEDF plays a crucial role in determining plasma-surface interactions and directly influences ion-assisted processes such as semiconductor etching and thin-film deposition \cite{taccogna2016non,seong2022development}.
Experimentally, EEDFs are commonly reconstructed from Langmuir probe measurements using the Druyvesteyn method, while IEDFs are typically measured using retarding field energy analyzers or energy resolved mass spectrometers\cite{eedf1, druyvesteyn1930niedervoltbogen, iedf1}. Although these techniques provide valuable kinetic information, they are often invasive, spatially localized, or experimentally demanding \cite{benedikt2021foundations}. Recent advances in non-invasive optical diagnostics, including optical emission spectroscopy (OES),  phase resolved OES (PROES), laser induced fluorescence (LIF), and absorption spectroscopy, have enabled spatially and temporally resolved measurements of several macroscopic plasma observables\cite{hanna2020investigating,engeln2020foundations, schulze2007space,beckfeld2025fiber}. Combined with appropriate spectroscopic or collisional-radiative models, these techniques can provide valuable information on plasma density, temperature, species concentration, and electric field distributions, thereby offering increasingly rich macroscopic descriptions of plasma behavior.
In many plasma processing applications, the spatial distributions of electron and ion energies govern the local etch and deposition rates, making spatially resolved kinetic information essential for achieving process uniformity and improving device performance \cite{lim2021wafer, sharma2014spatially, han2023wafer}. Furthermore, reactor geometry, sheath dynamics, and non-uniform power deposition can produce significant spatial variations in the underlying energy distributions \cite{park2018evolution}, making bulk or spatially averaged measurements insufficient for accurately characterizing plasma behavior.
Despite significant advances in plasma diagnostics, obtaining spatially resolved EEDFs and IEDFs remains a challenging task. Many existing non-invasive approaches estimate EDFs indirectly by relying on assumed functional forms or inversion techniques, whose accuracy may deteriorate under strongly non-equilibrium plasma conditions\cite{schulze2007space}. At the same time, advanced diagnostics have enabled spatially resolved measurements or estimates of several macroscopic plasma quantities, including plasma density and temperature, depending on the diagnostic technique and analysis methodology employed \cite{kim2013two,barnat2010two, weatherford2012two}. These developments naturally raise the question of whether such readily measurable macroscopic plasma observables contain sufficient information to reconstruct the spatially resolved energy distribution of the underlying kinetic state of the plasma, thereby motivating the inverse problem investigated in the present work.

Kinetic simulations such as PIC provide access to both macroscopic plasma quantities (e.g. density, temperature, and potential) and the underlying velocity or energy distributions\cite{birdshal,hybridpic, kim2005particle}. Consequently, large simulation datasets contain paired information describing both the macroscopic and microscopic state of the plasma.  
While kinetic simulations provide simultaneous access to both macroscopic plasma quantities and particle energy distributions, the extent to which macroscopic observables encode sufficient information to reconstruct the underlying kinetic state remains largely unexplored.
The relationship between macroscopic observables and kinetic distributions is highly nonlinear. Electron energy distributions are shaped by electric fields, collisions, plasma potential structures, and the high mobility of electrons, whereas ion energy distributions are additionally influenced by ion temperature, sheath acceleration, and ion-neutral collision processes\cite{kolobov2019electron,hippler2001low,boeuf2013rotating}. 
Consequently, reconstructing spatially resolved energy distributions from macroscopic information constitutes a challenging inverse problem.
The conventional EEDF and IEDF can subsequently be obtained by spatial marginalization of these distributions.
Rather than imposing predefined mathematical assumptions on the distribution shape, PIC simulations can be used to generate ground truth datasets (reference kinetic datasets) for training data-driven models to learn this inverse mapping \cite{arellano2024machine,mlphysicsreview}. 
In the present work, we distinguish between the conventional EDF, obtained after integrating over the spatial domain, and its spatially 
resolved representation. Specifically, the kinetic data are represented as energy distributions resolved simultaneously in space and energy. Spatial integration of this representation yields the corresponding EEDF or IEDF, while restricting the spatial domain allows distributions associated with specific regions, 
such as the sheath.

Data-driven tools and techniques have become an indispensable tool in plasma diagnostics \cite{anirudh20232022, ghosh2023deep, jetly2021extracting,arellano2024machine, kambara2023science, bentz2024project,kanarik2023human}. Training machine learning (ML) models directly on experimental diagnostic data can introduce measurement induced biases and is constrained by limited data availability. In contrast, PIC simulations self-consistently resolve particle dynamics and fields, providing access to the particle resolved phase-space information of charged species~\cite{garrigues2016appropriate}, thereby accurately providing both macroscopic and microscopic plasma parameters, and therefore creating an opportunity to investigate whether the latter can be inferred from the former\cite{arellano2024machine}.
Recent advances in deep learning have enabled the development of architectures capable of learning complex mappings between high-dimensional spatial fields. Convolution based models, graph based networks, and neural operator frameworks provide distinct inductive biases for representing physical systems and offer a promising avenue for addressing inverse problems in LTP physics \cite{mlltp, mlphysicsreview}.  

Although the mapping from kinetic distributions to macroscopic plasma quantities is well established through velocity-space moments \cite{alvesltpreview}, the inverse mapping is considerably more challenging because much of the microscopic information is compressed into a limited set of observable quantities. Demonstrating that a statistically meaningful approximation to this inverse mapping can nevertheless be learned from representative kinetic simulations would suggest that macroscopic plasma observables retain sufficient information about the underlying kinetic state. Motivated by this fundamental question, the present work investigates whether spatially resolved EEDFs and IEDFs can be reconstructed from macroscopic plasma quantities using data-driven learning models trained on high-fidelity PIC-MCC simulations data. Three representative deep learning models corresponding to different learning paradigms, namely convolutional encoder-decoder U-Net, Fourier Neural Operator (FNO), and graph based MeshGraphNet, are systematically evaluated for learning this nonlinear macro-to-micro mapping. Beyond image based reconstruction accuracy, the predicted distributions are assessed using physics based validation to determine whether the recovered kinetic information remains physically consistent. The results provide a proof of concept that nonlinear mappings between macroscopic observables and kinetic distributions can be learned from representative plasma states, with potential implications for reduced order plasma modeling, data-driven plasma simulations, surrogate plasma simulations, ML aided plasma diagnostics, and in the longer term, important building block for future physics informed digital twins for LTP applications\cite{Trieschmann,oehrlein2024future,digitaltwin}.


\section{Data-driven Methodology}\label{Data-driven Methodology}
\begin{figure}[h]
\centering
\includegraphics[width=0.7\linewidth]{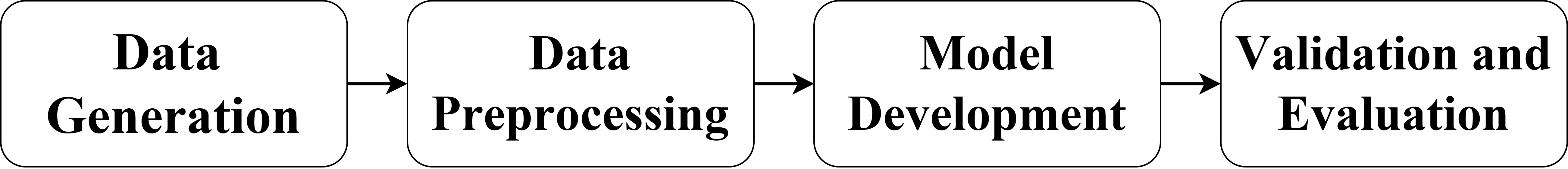}
\caption{General workflow followed in this study, proceeding from data generation and preprocessing to model development, followed by validation and evaluation.}
\label{fig:abstract_workflow}
\end{figure}

To investigate whether macroscopic plasma observables contain sufficient information to reconstruct the underlying kinetic state, we now describe a data-driven framework for learning the nonlinear mapping between spatially resolved plasma quantities and the corresponding spatially resolved energy distributions. The complete workflow consists of four sequential stages, illustrated in Fig.~\ref{fig:abstract_workflow}.
 First, paired datasets comprising spatially resolved macroscopic plasma observables and the corresponding EEDFs/IEDFs are generated using high-fidelity PIC-MCC simulations. Although the present study relies on simulation data, the proposed framework is sufficiently general to be extended to experimentally acquired datasets. Second, the data are preprocessed, normalized, and organized into multi-channel spatial representations suitable for deep learning (DL). Third, representative DL architectures are trained to learn the nonlinear mapping. Finally, the reconstructed EEDFs and IEDFs are evaluated using both image based performance metrics and physics-based validation to assess their reconstruction accuracy and physical consistency.


Within this framework, the inverse problem is formulated as a supervised learning task in which spatially distributed macroscopic plasma quantities are mapped to the corresponding spatially resolved energy distributions. Let $\mathcal{X}$ denote the space of admissible macroscopic states and $\mathcal{Y}$ denote the space of corresponding microscopic energy distributions. Each element \(X \in \mathcal{X}\) consists of macroscopic quantities defined over a spatial domain, while each \(Y \in \mathcal{Y}\) represents the associated discretized distribution function. In practice, both \(X\) and \(Y\) are discretized as multi-channel two-dimensional tensors defined on a common spatial grid. The underlying physical relationship between these quantities can be expressed as an unknown mapping $f^* : \mathcal{X} \rightarrow \mathcal{Y}$, which is highly nonlinear and analytically intractable. The goal of this work is to approximate this mapping using a neural network parameterized by $\theta$, denoted as $f_\theta$, such that $\hat{Y} = f_\theta(X)$. Given a dataset $\{(X_i, Y_i)\}_{i=1}^M$ consisting of paired samples obtained from 2D PIC-MCC simulations, the optimal parameters \(\theta^*\) are estimated by minimizing a loss functional $\mathcal{L}$ that measures the discrepancy between predicted and reference distributions:

\begin{equation}
\theta^* = \arg\min_\theta \frac{1}{M} \sum_{i=1}^M \mathcal{L}\big(f_\theta(X_i), Y_i\big).
\label{eq:optimal_param}
\end{equation}

This formulation captures the inverse problem of reconstructing microscopic distributions from macroscopic observables in a data-driven manner. Unlike traditional analytical approaches, which often rely on simplifying assumptions that may not hold across different regimes, the proposed approach leverages statistical learning to approximate the underlying mapping directly from data. 
 

\section{Data Generation}
\label{data generation}
This section describes the generation of the datasets, used throughout this work, which provide a controlled reference against which the reconstruction can be evaluated. We first provide a brief overview of the 2D PIC-MCC methodology, followed by the simulation setup, dataset generation procedure, and construction of the input-output pairs used for model training and evaluation.

\subsection{Particle-in-Cell (PIC) Overview} \label{sec:pic_overview}
The PIC-MCC method self-consistently couples charged-particle dynamics with the evolution of the electrostatic fields through an iterative cycle \cite{hybridpic, garrigues2016appropriate,shah2017novel}. The simulation is initialized by specifying the 2D computational grid, charged particle distributions, simulation time step, and boundary conditions. During each time step, particle charges are deposited onto the computational grid using a charge-assignment scheme to obtain the spatial charge density. The electrostatic potential is then computed by solving Poisson's equation, from which the electric field is evaluated on the grid. These fields are interpolated back to the particle locations, and the particle trajectories are advanced using the Boris particle pusher. Particle collisions with the background gas are subsequently treated statistically using the Monte Carlo Collision (MCC) method, after which the cycle is repeated until the desired simulation time is reached \cite{garrigues2016appropriate, hybridpic}. The dataset for this study has been generated using our in-house, massively parallel 2D-3V GICS-PIC code, which has been rigorously benchmarked and verified against established reference cases \cite{sharma2025electron,varghese2025benchmarkingparallelizationelectrostaticparticleincell, charoy20192d, parodi2025step, varghese2026insights}. In this code, particle phase-space data are stored in an array of structures (AoS) referred as \textit{particle} data structure, while grid-based quantities such as density, temperature, potential, and field components are discretized on a two-dimensional mesh and stored in one-dimensional arrays for computational efficiency\cite{hybridpic}.

\subsection{Simulation Setups for Data Generation} \label{sec:sim_params_and_data_gen}

\begin{table}[]
\centering
\caption{Simulation cases and corresponding physical conditions.}
\label{tab:testcases}

\begin{tabular}{|c|p{3.0cm}|c|c|p{.0cm}|}
\hline
\textbf{Case} &
\textbf{Description} &
\textbf{E field} &
\textbf{B field} &
\textbf{MCC collisions} \\
\hline
\hline
1 &
Electric field only &
Yes &
$B_z=0$ &
No \\
\hline
2 &
High magnetic field &
Yes &
$B_z=2\times10^{-4}$ T &
No \\
\hline
3 &
Collisions, low pressure &
Yes &
$B_z=0$ &
Yes ($n_g=3\times10^{19}\,\mathrm{m^{-3}}$) \\
\hline
4 &
Collisions, high pressure &
Yes &
$B_z=0$ &
Yes ($n_g=3\times10^{20}\,\mathrm{m^{-3}}$) \\
\hline
5 &
Ionization, high pressure &
Yes &
$B_z=0$ &
Yes + ionization ($E_{\mathrm{iz}}=1.43$ eV) \\
\hline
6 &
Low magnetic field with collisions, low pressure &
Yes &
$B_z=2\times10^{-5}$ T &
Yes ($n_g=3\times10^{19}\,\mathrm{m^{-3}}$) \\
\hline
7 &
High magnetic field with collisions, low pressure &
Yes &
$B_z=2\times10^{-4}$ T &
Yes ($n_g=3\times10^{19}\,\mathrm{m^{-3}}$) \\
\hline
\end{tabular}
\end{table}

\begin{figure}[]
    \centering
\includegraphics[width=0.95\linewidth]{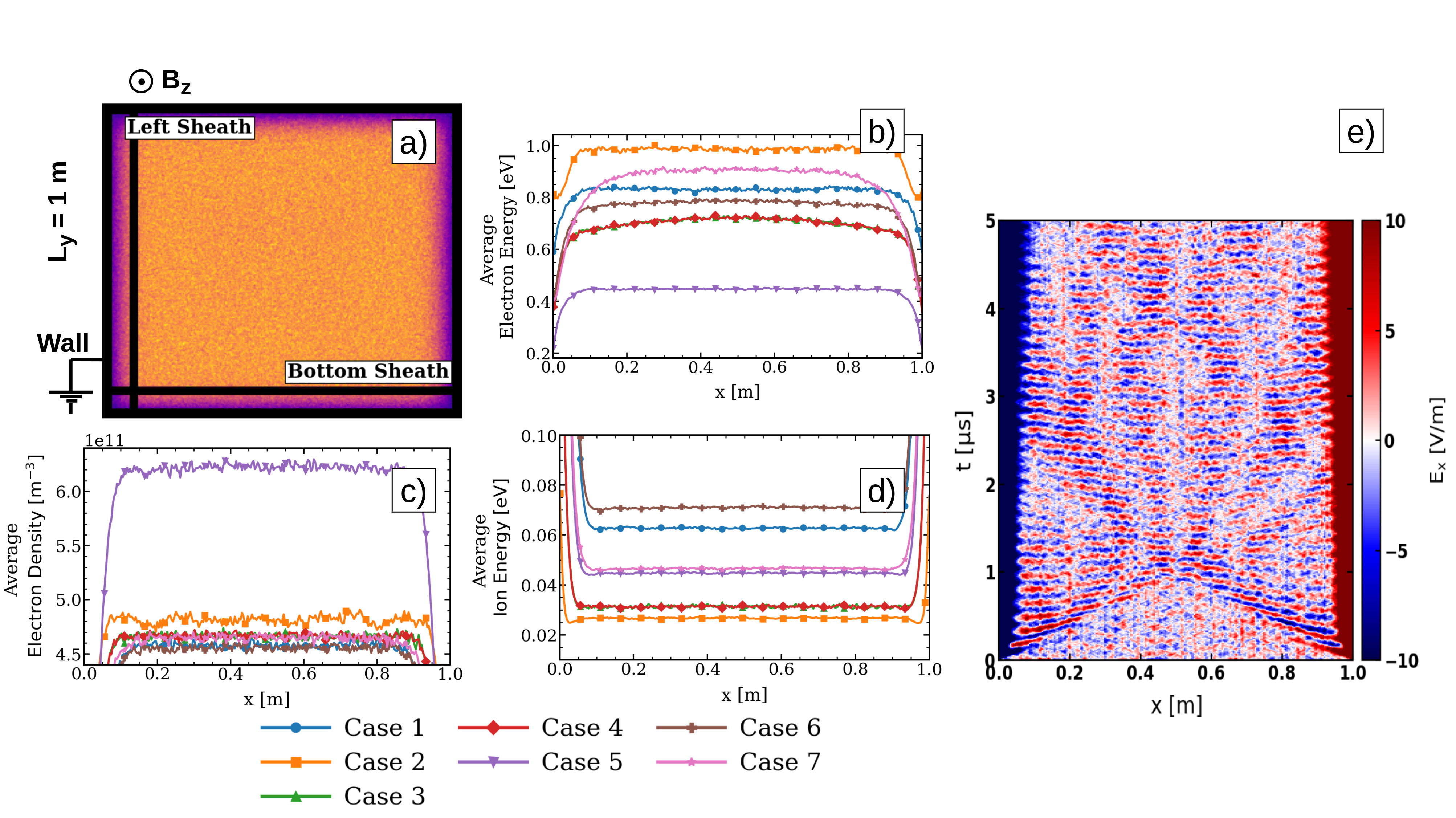}
    \caption{a) Simulation setup for case 1 at $5 \mu s$ showing plasma density. b) Electron energy plots for various cases averaged along the $y$ direction. c) Electron density plots for various cases averaged along the $y$ direction. d) Ion energy plots averaged along the \textit{y} direction. e) Space time evolution of $X$ component of electric field exactly at the middle of the domain showcasing the evolution of simulation through the transient phase (case 1). }
    \label{fig:SimDom}
\end{figure}

The dataset used in this work has been generated from a series of simulations, following the step by step verification framework as proposed by Parodi and Petronio ~\cite{parodi2025step}. The goal of performing multiple simulations using different test cases is to incrementally add physics-based complexity to the problem, providing diverse data for model training. The common simulation parameters across all test cases are as follows. The computational domain is a $1 \times 1$~m square, with grounded walls and potential set to zero on all four sides, making it a conductive wall that absorbs all particles. There are 256 cell in each direction making $\Delta x = 3.9 \times 10^{-3}$, the simulation domain is initialized with a uniform distribution of quasineutral hydrogen plasma with temperatures from a Maxwell-Boltzmann distribution ($n_e = n_i = 5 \times 10^{11}~\mathrm{m^{-3}}$, $T_e = 1$~eV, $T_i = 300$~K), and the simulation evolved over $t_f = 5~\mu$s with a time step $\Delta t = 2.5 \times 10^{-9}$~s. The simulations are performed using 5 million macroparticles. The physical significance of the generated data lies in its ability to capture electron and ion density evolution, sheath formation, collisional damping, and ionization driven density growth under controlled conditions. The dataset comprises around 2000 frames per simulation, yielding thousands of samples. The computational costs remained modest, making the dataset scalable for both benchmarking and ML applications. All quantities are stored as double precision arrays. Spatially varying macroscopic plasma quantities are calculated on the uniform computational grid, while the corresponding electron and ion energy distributions are stored as discretized histograms. Details of physical configurations are shown in Table \ref{tab:testcases}.

\begin{figure}[]
    \centering
\includegraphics[width=0.6\linewidth]{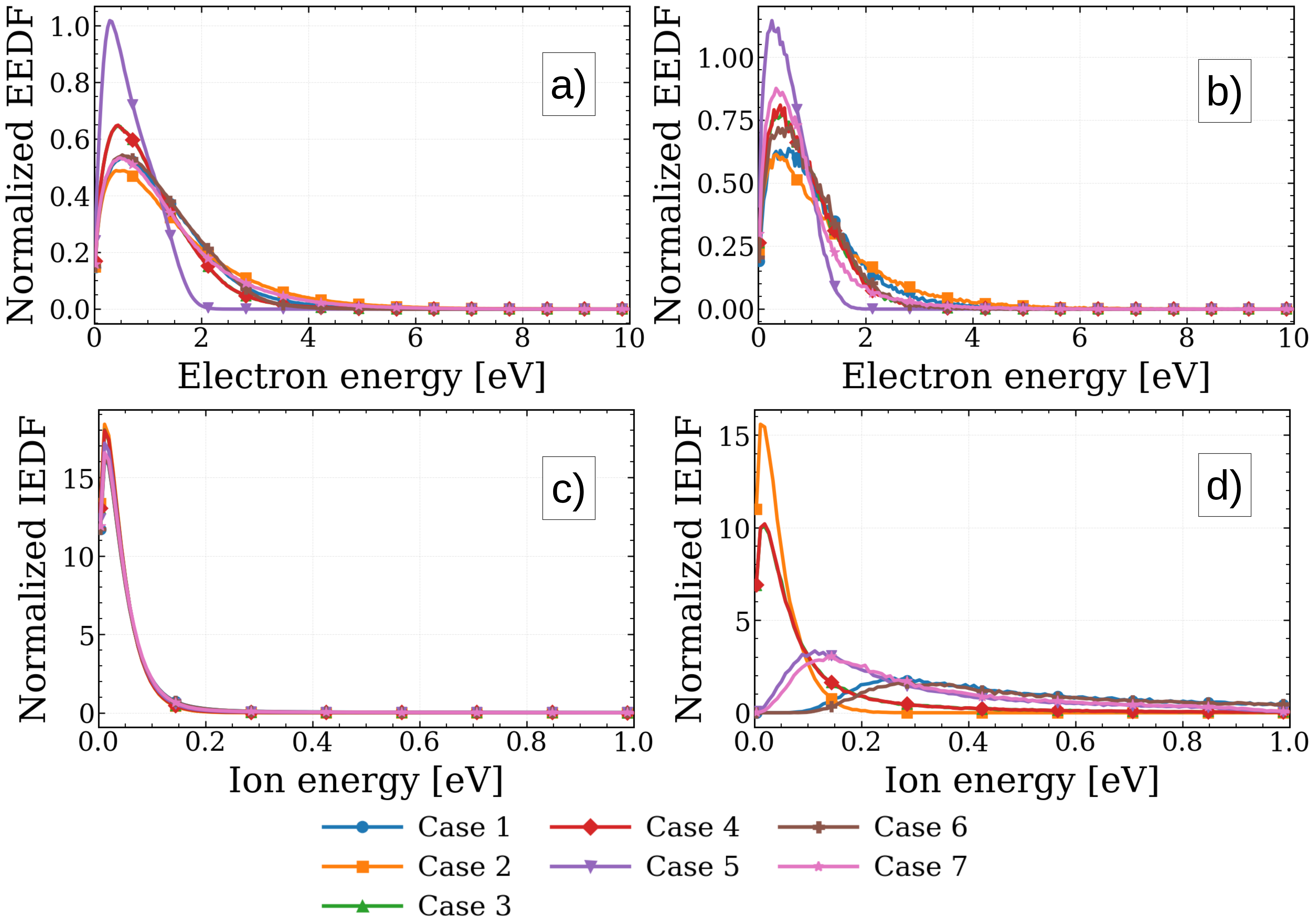}
    \caption{a) EEDF for full simulation domain. b) EEDF for left sheath.
    c) IEDF for full simulation domain.
    d) IEDF for left sheath. (All the EDFs are calculated at time = $5\mu s$)}
    \label{fig:edf_output}
\end{figure}

Physically, the selected cases represent distinct kinetic characteristic of the plasma. In the electric field only case (Table \ref{tab:testcases}, case 1), the faster loss of electrons to the absorbing walls produces a space-charge sheath that separates the quasineutral bulk from the wall, which can be seen in Fig. \ref{fig:SimDom}, this results in accelerated ions in the sheath as shown in Fig. \ref{fig:edf_output} d).  Consequently, the continued loss of plasma causes the sheath region to progressively expand with time, as shown in Fig. \ref{fig:SimDom} e).
When a magnetic field is applied (Table \ref{tab:testcases}, case 2), the particle motion is governed by the cyclotron frequency and Larmor radius, thereby confining electrons and reducing wall losses, resulting in a smaller sheath region as shown in Fig. \ref{fig:SimDom}, and a reduction in ion energies as shown in Fig. \ref{fig:edf_output} d). For cases 3-7 in Table \ref{tab:testcases}, a MCC model based on the algorithm of Vahedi et al\cite{vahedi1995monte} is used. For case with collisions in low pressure and high pressure (Table \ref{tab:testcases}  case 3 and 4) electron neutral collision includes elastic, charge exchange, excitation scattering cross section, and the data is taken from the Morgan database on LXCat\cite{pitchford2017lxcat}, while the ion neutral elastic and charge exchange collision cross section is based on the tabulated data of Schultz et al\cite{schultz2023data}. When MCC is included, the neutral gas density affects the rate of electron energy loss; therefore, the low and high pressure cases differ in the extent of momentum and energy relaxation, eventually affecting the space charge separations, see Fig. \ref{fig:SimDom} and ion energies in the sheath see Fig. \ref{fig:edf_output} d). Electron impact ionization is included using the data from  Janev et al.,\cite{janev1987elementary} with the ionization threshold scaled to 1.43 eV (Table \ref{tab:testcases}  case 5), which introduces a particle source and causes the rate of change in density and particle loss at the walls, see Fig. \ref{fig:SimDom}. By combining a magnetic field with collisions, two more physically distinct cases are considered (Table \ref{tab:testcases}, cases 6 and 7), in which both magnetic confinement and collisional energy loss are accounted for. Overall, Fig. \ref{fig:edf_output} demonstrates that distinct kinetic regimes exist within the sheath for different cases; however, when the energy distribution is evaluated over the entire computational domain, the features can be averaged out, thereby masking the localized dynamics near the boundary. This physics diverse macro-micro dataset provides a controlled benchmark for evaluating whether the learned inverse mapping generalizes across changes in the underlying plasma conditions.

\subsection{Input and Output Structure of the PIC Simulation Data}
\label{sec:pic_ip_op}
The raw outputs from PIC-MCC simulations are organized into structured input-output pairs that capture the relationship between measurable macroscopic quantities and corresponding microscopic energy distributions. The input data is defined on a uniform Cartesian grid of size $N_y \times N_x$,  where $N_y$ is the number of cells in the $y$ direction and $N_x$ is the number of cells in the $x$ direction, capturing spatial variations of macroscopic plasma properties across the computational domain. The input space comprises 3 macroscopic parameters extracted from the PIC simulation: electron density ($n_e$), electron energy ($T_e$), and ion energy ($T_i$). Each parameter is represented as a two-dimensional profile on the computational mesh. The corresponding output space is derived from the particle energy histograms based on the phase-space information stored in the \textit{particle} data structure. These histograms yield spatially resolved electron and ion energy distributions, which are used to compute the EEDF and IEDF. Specifically, four output profiles are considered: $\epsilon_{ex}, \epsilon_{ey}, \epsilon_{ix}, \epsilon_{iy}$, where the first subscript ($e$ or $i$) denotes the particle species, and the second subscript ($x$ or $y$) denotes the spatial axis along which the distribution is resolved. These normalized histograms form the target quantities for model training.

The computational domain, a square bounded by $(X_{\min}, Y_{\min})$ and $(X_{\max}, Y_{\max})$, is discretized into $N_y \times N_x$ cells of uniform size, with dimensions
\begin{equation}
\Delta x = \frac{|X_{\max} - X_{\min}|}{N_x}, 
\qquad 
\Delta y = \frac{|Y_{\max} - Y_{\min}|}{N_y}.
\label{eq:ip_bin_width}
\end{equation}

\begin{figure}[]
    \centering
    \includegraphics[width=0.9\linewidth]{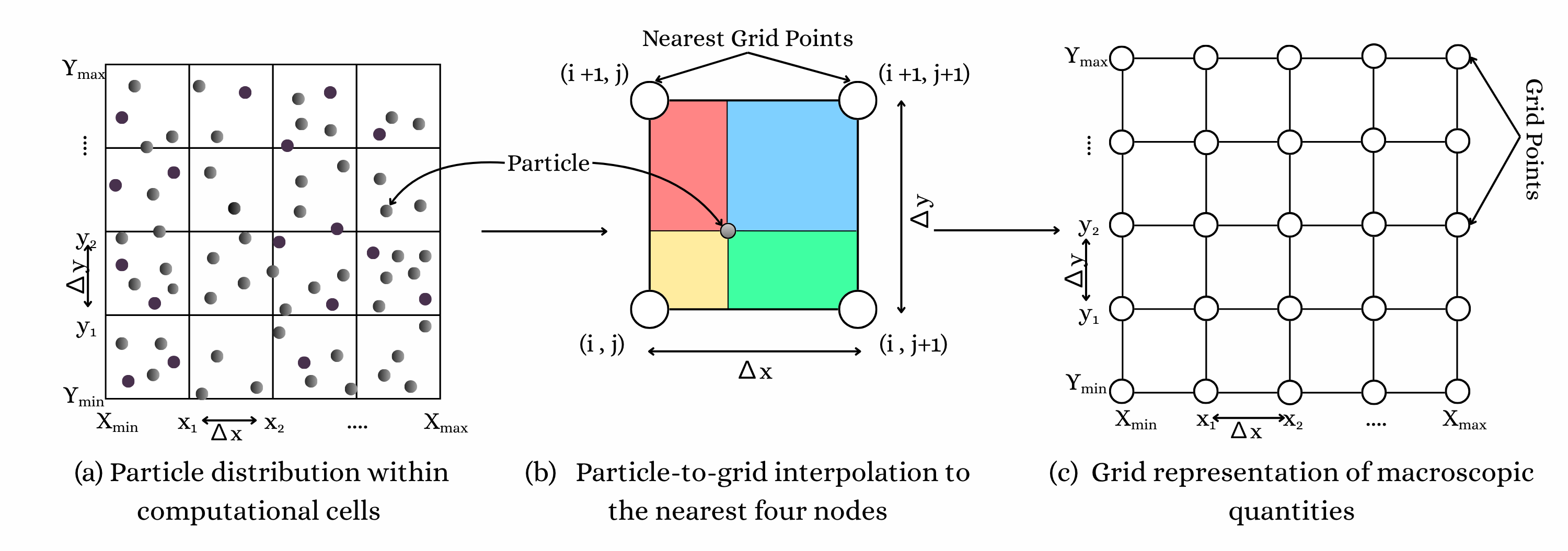}
    \caption{Schematic representation of dataset construction from two-dimensional PIC simulations.
    (a) The computational domain is divided into 
    $N_y\times N_x$ cells of size $\Delta x$ and $\Delta y$, with multiple simulation particles distributed in each cell.
    (b) Each particle contributes to its four nearest grid nodes according to its position within the cell, ensuring smooth spatial deposition of physical quantities
    (c) The resulting node centered mesh ($G_y\times G_x$), where macroscopic fields such as densities and temperatures are extracted for use as input data for training.}
    \label{fig:macro_grid}
\end{figure}

\noindent

Each cell is indexed by $(i,j)$ and spans the region $(Y_i, Y_i+\Delta y) \times (X_j, X_j+\Delta x)$. Inside every cell, many simulation particles with different velocities are present. Consider one such particle with coordinates $(x_p, y_p)$ within the cell. Its contribution is distributed to the four nearest grid vertices (the cell corners) rather than being assigned to a single location. This interpolation follows the PIC scheme, where weight of each particle  is distributed among the four nearest vertices in proportion to the rectangular sub-areas formed between the particle position and the diagonally opposite grid vertex. In this way, each particle contributes partially to multiple vertices. Thus, while the computational space consists of $N_y \times N_x$ cells, the macroscopic fields are stored on $(N_y + 1) \times (N_x + 1)$ grid vertices. For simplicity, we denote $G_y = N_y + 1$ and $G_x = N_x + 1$, so the final matrices representing the macroscopic quantities are of size $G_y \times G_x$. As illustrated in Fig.~\ref{fig:macro_grid}, each grid entry $(i,j)$ corresponds to the average value of the physical quantity over its surrounding spatial region. For example, the electron temperature $T_e$ at grid point $(i,j)$ denotes the mean kinetic energy of all electrons in the associated spatial volume, rather than a pointwise measurement. This averaging is intrinsic to the PIC methodology and is crucial for interpreting the macroscopic data in a physically consistent manner.

\begin{figure}[]
    \centering
    \includegraphics[width=0.95 \linewidth]{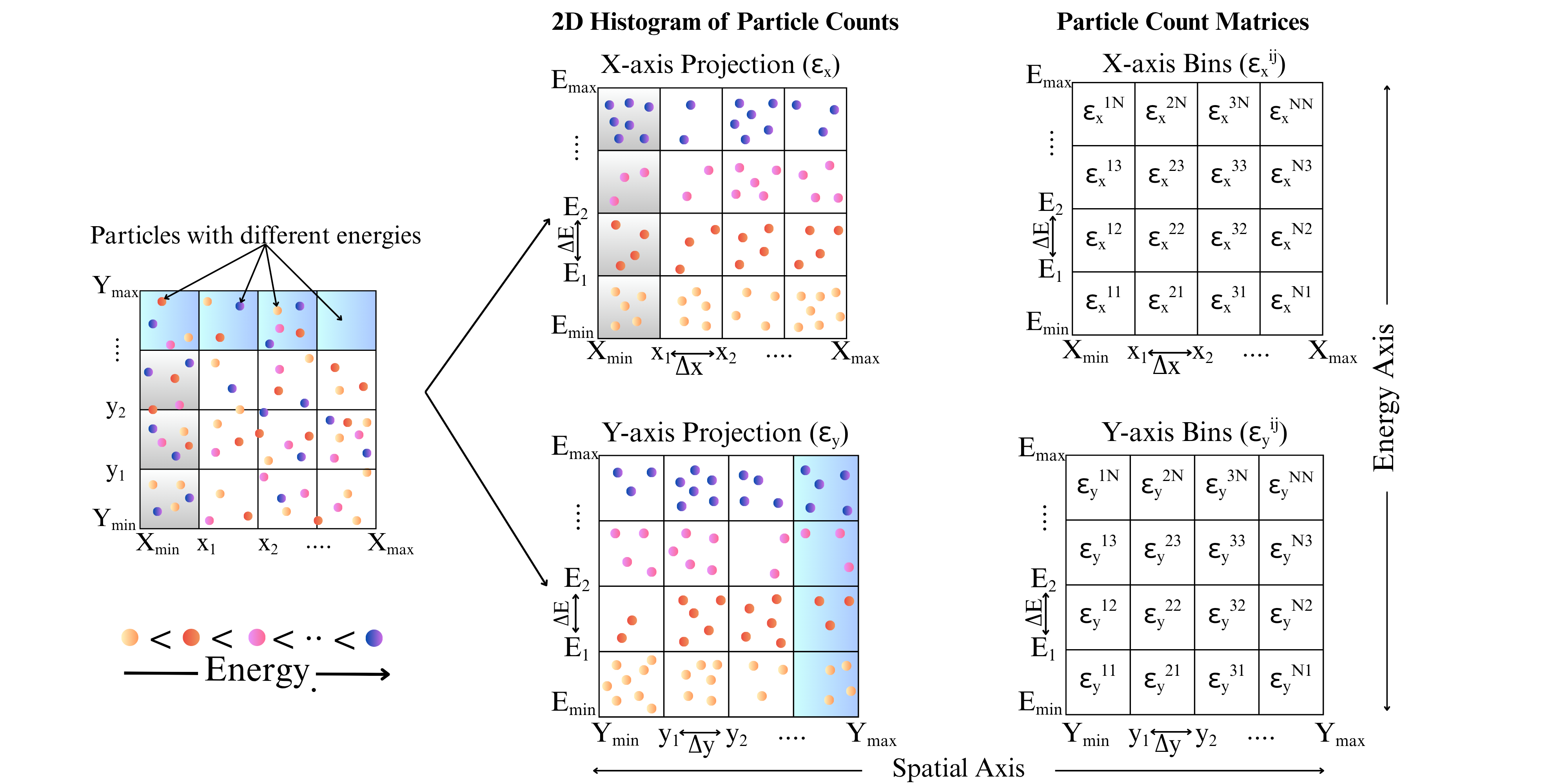}
    \caption{Schematic representation of the output space as 2D histograms of particle energy discretized in space. In these histograms the vertical axis corresponds to energy bins $(E_1, E_2, \dots, E_{\max})$ and the horizontal axis corresponds to spatial discretization. Four such histograms are constructed, two for electrons and two for ions, where one set resolves the distribution along the $x$-axis and the other set along the $y$-axis. (a) Particles with different energies are distributed across a two-dimensional spatial grid. (b) This illustrates the 2D histograms $\epsilon_x$ and $\epsilon_y$, where particles are grouped into discrete space-energy bins. (c) The corresponding particle count matrices $\epsilon^{ij}_{x}$ and $ \epsilon^{ij}_{y}$ represent the 2D histograms, where each element denotes the number of particles in spatial cell $i$ and energy bin $j$.}
    \label{fig:micro_grid}
\end{figure}

The output representation is shown in Fig ~\ref{fig:micro_grid}. Each profile is organized as an $M_e \times M_s$ matrix, where the horizontal axis corresponds to the spatial discretization of the computational domain and the vertical axis represents the discretized energy space. The energy axis spans the interval $[E_{\min}, E_{\max}]$, with resolution $\Delta E$, while the spatial axis spans $[S_{\min}, S_{\max}]$ with resolution $\Delta s$. These resolutions are chosen based on relevant physical considerations. In this work, the spatial resolution matches that of the input fields, specifically, when projecting particle energies onto the $x$- and $y$-axes, $\Delta s$ is set equal to $\Delta x$ and $\Delta y$, as defined in Eq. ~\ref{eq:ip_bin_width}.

\begin{equation}
\Delta E = \frac{|E_{\max} - E_{\min}|}{M_e}, 
\qquad 
\Delta s = \frac{|S_{\max} - S_{\min}|}{M_s}.
\label{eq:op_bin_width}
\end{equation}

Each matrix entry $(i,j)$ denotes the number of particles contained in the spatial strip $(s_j, s_j + \Delta s)$ whose energies fall within the interval $(E_i, E_i + \Delta E)$. This formulation yields a spatially resolved description of the particle energy distributions, forming the basis of the supervised learning task addressed in this work. The main motivation for choosing this representation as the model output is that it provides a more detailed view of the spatial variations in the particle energy distribution. Unlike the input space, where each grid vertex stores values obtained by summing the contributions of many particles, thereby averaging out the information and losing fine details, these two-dimensional histograms preserve the actual particle distributions along with their corresponding energy ranges.

The two-dimensional space energy histograms must be transformed into a physically meaningful energy distribution function (EDF). The normalized EDF is obtained by summing the histogram counts along the spatial axis, thereby marginalizing over spatial coordinates. Let the histogram be denoted by $\epsilon_{ij}$, representing the particle count in the bin corresponding to energy interval $(E_i, E_i + \Delta E)$ and spatial interval $(x_j, x_j + \Delta x)$. The normalized energy distribution is computed as follows.

\begin{equation}
    f(E_i \leq \epsilon \leq E_i + \Delta E) = \frac{1}{\Delta E} \cdot 
    \frac{\sum_{j=0}^{N-1} \epsilon^{ij}}
    {\sum_{k=0}^{M-1}\sum_{j=0}^{N-1} \epsilon^{kj}}
\label{eq:hist_to_dis}
\end{equation}

This transformation yields a normalized one-dimensional EDF that is directly comparable with experimentally relevant plasma energy distributions. For the full domain EDF, the summation in Eq. \ref{eq:hist_to_dis} is carried out over either of the complete $\epsilon_{x,y}$ two-dimensional space energy histograms. In contrast, for the left sheath region shown in Fig. \ref{fig:SimDom} a), the \textit{x} resolved space energy histogram, denoted by $\epsilon_{ex}$, is used and the spatial summation is restricted only to the first \textit{k} spatial columns adjacent to the left boundary, rather than over the full spatial extent. Similarly, to obtain the EDF for the bottom sheath region, the \textit{y} resolved space energy histogram, denoted by $\epsilon_{ey}$, is used, and the summation is restricted to the first \textit{k} spatial columns which corresponds to the bottom boundary. The sheath index $k$ is determined using the Brinkmann integral criterion \cite{brinkmann2007beyond, schulze2007space}, as given in Eq. \ref{eq: sheath}. Here, $x=0$ denotes the wall position, and $x$ is measured from the wall toward the plasma. The sheath edge is therefore located at $x=k\Delta x$. The quantities $n_e(x)$ and $n_i(x)$ represent the electron and ion density profiles, respectively, averaged along the direction parallel to the wall.
\begin{equation}
\int_{0}^{k\Delta x} n_e(x),dx
=
\int_{k\Delta x}^{\infty}
\left[n_i(x)-n_e(x)\right]dx 
\label{eq: sheath}
\end{equation}
Using Eq. \ref{eq: sheath}, and taking into account the spatial variation of the electrostatic potential and plasma density near the boundary for the present discharge conditions, an average value of $k=10$ is used across all cases for sheath specific analysis. Thus, the same normalization procedure is applied to each region, but the spatial summation limits are modified according to the region of interest, and the  EDF for the sheath is generated as shown in Fig. \ref{fig:edf_output}. In this work, the term bulk distribution refers to the distribution obtained over the complete simulation domain, this approximation is adopted because the 
sheath occupies only a relatively small fraction of the domain.


\section{Data Preprocessing}
\label{sec:data_preprocessing}

\begin{figure}[]
    \centering
    \includegraphics[width=0.6\linewidth]{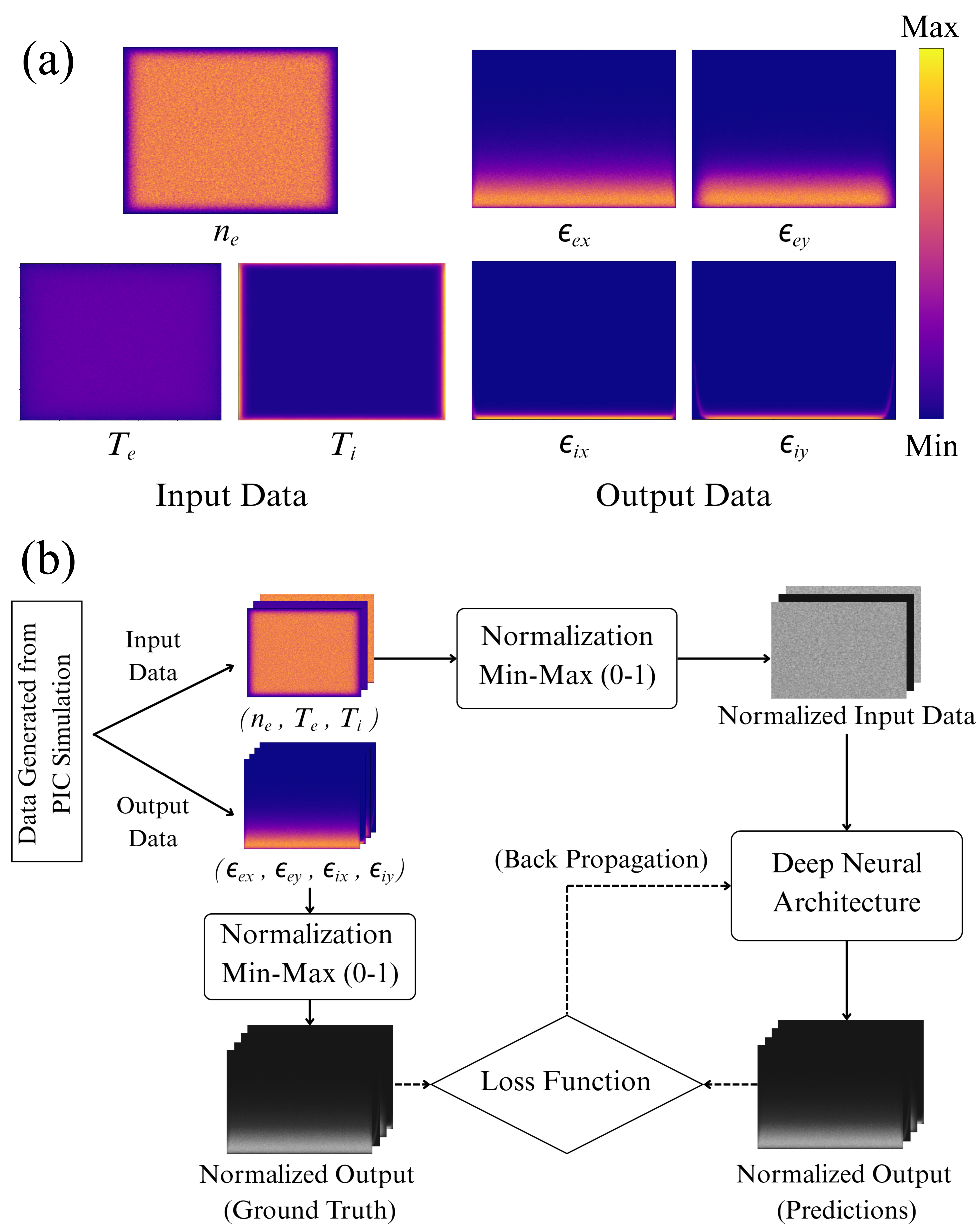}
    \caption{Overview of the proposed data-driven framework. (a) Representative input macroscopic observables and corresponding output microscopic  energy distribution functions. (b) Workflow illustrating the data preprocessing, neural network training, prediction, and performance evaluation stages adopted in this study.}
    \label{fig:pipeline}
\end{figure}

 The overall DL based framework, including data preprocessing, model training, and evaluation, is illustrated in Fig.~\ref{fig:pipeline}. In addition to normalization, we analyze the distribution of the data, across different simulation cases, to verify that the dataset spans diverse physical regimes.

\subsection{Data Distribution Analysis}
\begin{figure}
    \centering
    \includegraphics[width=0.85\linewidth]{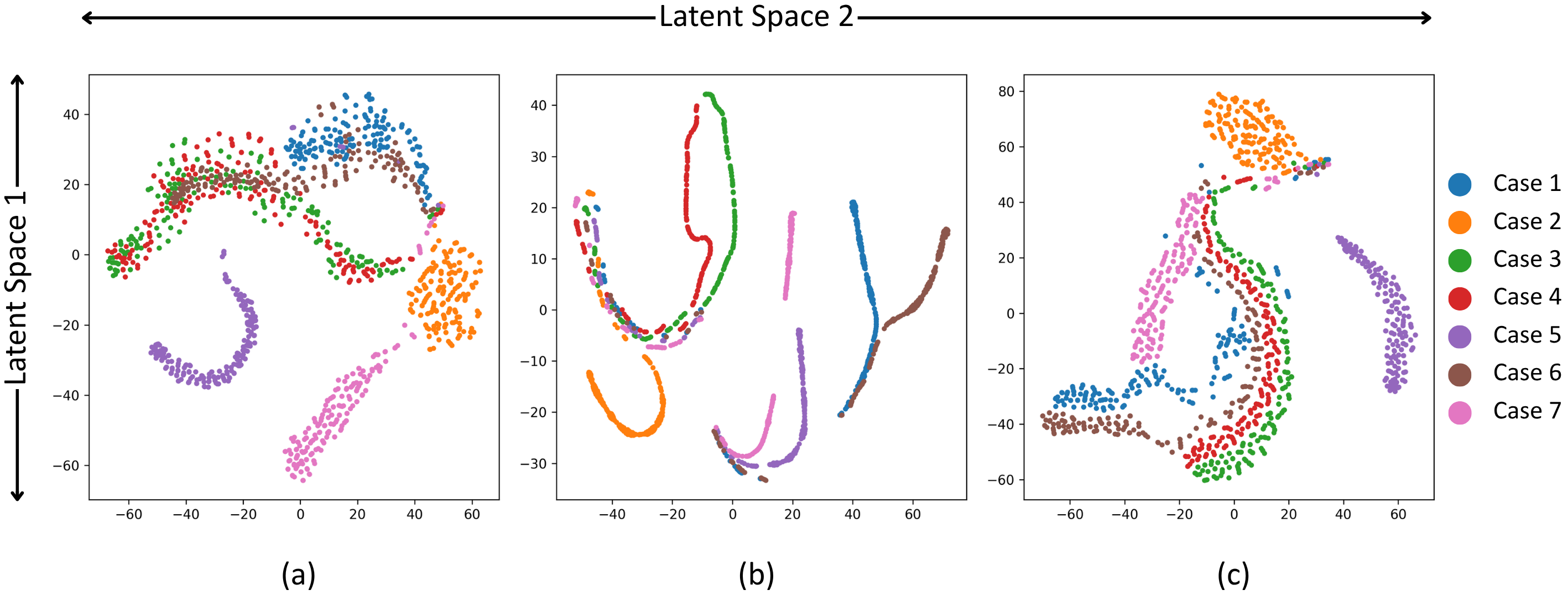}
    \caption{t-SNE visualization of input channels: (a) electron energy, (b) ion energy, and (c) electron density. Each point represents a data sample projected into a two-dimensional embedding space using t-SNE with perplexity 30. Points are colored by simulation case, showing clear clustering and indicating the presence of distinct physical regimes in the dataset.}
\label{fig:tsne_visualization}
\end{figure}
Prior to model training, we analyze the dataset to ensure that different simulation cases exhibit distinct data distributions. Since the dataset is obtained from multiple PIC-MCC simulation cases corresponding to different physical operating conditions (Table~\ref{tab:testcases}), it is important to confirm that the cases are sufficiently distinct, as this would reduce the significance of cross-regime evaluation. To assess the distributional differences, 200 samples are randomly selected from each simulation case. These samples are normalized using the channel-wise min-max scaling described in Section~\ref{sec:normalization} to ensure consistent representation across cases. We then apply t-distributed stochastic neighbor embedding (t-SNE) \cite{JMLR:v9:vandermaaten08a} to selected macroscopic input channels. For each channel, the high-dimensional spatial fields are embedded into a two-dimensional latent space using a perplexity of 30. The resulting embeddings, visualized for all samples and color labeled by simulation case, reveal clear clustering patterns corresponding to different data cases as shown in Fig. \ref{fig:tsne_visualization}. These observations provide a qualitative visualization of differences among the simulation cases, supporting the adopted training-testing strategy, where entire simulation cases can be held out for testing.

\subsection{Channel-wise Normalization}
\label{sec:normalization}
Macroscopic plasma quantities and the corresponding energy distribution histograms exhibit significant variation in magnitude across different channels and simulation cases, often spanning several orders of magnitude. Such differences in scale can lead to poorly conditioned optimization and bias the learning process toward higher-magnitude features, thereby requiring normalization \cite{sola1997importance}. To mitigate this effect, a channel-wise min-max normalization scheme is applied independently to both the input and output data. Let $c$ denote the channel index. Each input field, corresponding to macroscopic quantities, is represented as a two-dimensional array $X^{(c)} \in \mathbb{R}^{N \times N}$. The normalized input is computed as

\begin{equation}
X_{i,j}^{\prime (c)} = \frac{X_{i,j}^{(c)} - X_{\min}^{(c)}}{X_{\max}^{(c)} - X_{\min}^{(c)}},
\label{eq:ip_norm}
\end{equation}

where $X_{\min}^{(c)}$ and $X_{\max}^{(c)}$ denote the minimum and maximum values of the channel computed over the entire dataset. Similarly, the output channels, representing spatially resolved energy histograms, are normalized as

\begin{equation}
Y_{i,j}^{\prime (c)} = \frac{Y_{i,j}^{(c)} - Y_{\min}^{(c)}}{Y_{\max}^{(c)} - Y_{\min}^{(c)}},
\label{eq:op_norm}
\end{equation}

where $Y_{\min}^{(c)}$ and $Y_{\max}^{(c)}$ are defined analogously.

This normalization scales all input and output channels to the range $[0,1]$, ensuring that each physical field contributes proportionally during training while preserving its internal spatial structure. It also aligns the data with the sigmoid activation used in the output layer of all models, which constrains predictions to the same range. During inference, the predicted outputs are rescaled to their original physical ranges using the stored channel-specific normalization parameters, thereby maintaining consistency with the underlying simulation data.

In contrast to studies that focus solely on EEDFs or IEDFs\cite{kim2013two, han2023wafer}, the present work considers the simultaneous reconstruction of both EEDFs and IEDFs. This is a significantly more challenging problem because electrons and ions exhibit fundamentally different transport characteristics and respond to plasma conditions differently as mentioned in Section \ref{sec:intro}. Consequently, successful joint reconstruction requires the model to learn species specific kinetic behavior, making it a more comprehensive test of the proposed macro-to-micro mapping framework.

\section{Deep Learning Architectures for EDF Reconstruction}\label{sec:dl_model_implementation}
 To systematically study how different architectures, each characterized by a different representation paradigm, affect this learning task, we consider a convolutional architecture (U-Net), a neural operator-based model (Fourier Neural Operator), and a graph-based model (MeshGraphNet). U-Net, with its encoder-decoder structure and skip connections, is well-suited for learning localized spatial features with image-to-image mappings between finite dimensional Euclidean spaces \cite{ronneberger2015u}. The FNO \cite{li2020fourier} is included as an operator-based model that learns a mapping function between spaces rather than only between fixed-size arrays and, in principle, can be evaluated at resolutions different from those used during training. Finally, MeshGraphNet\cite{pfaff2020learning} is considered to investigate the role of graph-based representations. In plasma diagnostics experiments, complete dense plasma observable information may not always be available and it may be sparse. Graph-based models are naturally suited to such settings because they represent the system as nodes and edges rather than as a fixed Cartesian grid. This allows MeshGraphNet to learn spatial dependencies from sparse or irregular input data. Neural operator frameworks are particularly attractive for this problem because they are designed to learn mappings between fields rather than discrete variables, making them well suited for capturing the non-local relationships that arise in LTP scenarios.
 These models differ fundamentally in how they represent spatial information, capture dependencies, and generalize across resolutions.


\subsection{U-Net: Convolutional Encoder-Decoder Architecture}

\begin{figure}[]
    \centering
    \includegraphics[width=0.7\linewidth]{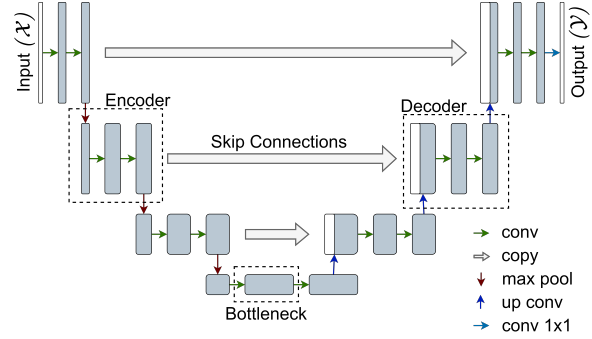}
    \caption{Schematic of the U-Net architecture used in this work. The input field $\mathcal{X}$ is processed through an encoder path consisting of repeated convolution and max-pooling operations to extract multi-scale feature representations. At the bottleneck, the latent features capture high-level contextual information. The decoder path reconstructs the target output through successive up-convolution and convolution operations. Skip connections concatenate feature maps from corresponding encoder and decoder levels, enabling the recovery of fine spatial details lost during downsampling. A final $1\times1$ convolution generates the predicted output field $\mathcal{Y}$.}
    \label{fig:unet}
\end{figure}

The U-Net architecture is a fully convolutional neural network originally developed for dense prediction tasks such as image segmentation~\cite{ronneberger2015u}. In the present work, both the macroscopic plasma quantities and the target energy distribution functions (EDFs) are defined on the same spatial grid, making U-Net a natural baseline for structured field-to-field regression \cite{desai2022deep}. Its convolutional design is particularly well suited for learning local spatial patterns while preserving fine-scale structure through multi-scale feature extraction \cite{lecun2015deep}. This is particularly relevant in plasma systems, where local measurements are influenced by neighboring regions through transport processes and field interactions\cite{mandikal2026benchmarking}. The architecture consists of a contracting path that progressively encodes the input into higher-dimensional feature representations, and an expanding path that reconstructs the output at the original spatial resolution as shown Figure \ref{fig:unet}. The encoder is composed of repeated convolutional blocks followed by downsampling operations, which increase the receptive field and enable the network to capture contextual information across larger spatial regions.

In our implementation, the U-Net operates on normalized macroscopic input fields 
while predicts the corresponding
energy distribution histograms. The network follows a four-level encoder-decoder design with a base feature width of 64 channels. At each downsampling stage, the encoder applies two successive $3 \times 3$ convolutional layers with ReLU activations, followed by a $2 \times 2$ max-pooling operation with stride 2. The number of feature channels is doubled at each level, enabling the network to learn increasingly rich hierarchical features as the spatial resolution decreases. At the bottleneck, the network processes the most compressed latent representation of the input through two additional convolutional layers with ReLU activation, allowing it to capture higher-level abstractions of the underlying plasma state. The decoder progressively reconstructs the output using successive $2 \times 2$ transposed convolution layers for upsampling. At each decoding stage, the upsampled features are concatenated with the corresponding encoder feature maps through skip connections. These skip connections are essential for retaining fine-scale spatial information that may otherwise be lost during downsampling, thereby enabling accurate reconstruction of localized features such as sharp gradients and sheath structures. Following each skip connection fusion, two additional $3 \times 3$ convolutional layers with ReLU activation are applied to refine the decoded features. A final $1 \times 1$ convolution layer maps the learned feature representation to the required output channels. A sigmoid activation function is used at the output layer to ensure that predictions remain within the normalized target range $[0,1]$, consistent with the adopted preprocessing strategy.

\subsection{Fourier Neural Operator: Spectral Neural Operator for Global Field Mapping}

\begin{figure}[]
    \centering
    \includegraphics[width=0.7\linewidth]{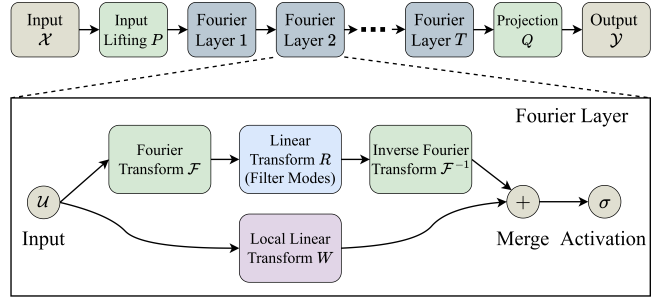}
    \caption{Architecture of the Fourier Neural Operator (FNO). The input field $\mathcal{X}$ is lifted to a higher-dimensional latent representation using the operator $P$ and processed through $T$ stacked Fourier layers. Each Fourier layer combines a spectral transformation, consisting of the Fourier transform $\mathcal{F}$, learnable spectral operator $R$, and inverse Fourier transform $\mathcal{F}^{-1}$, with a local linear transformation $W$. The outputs of the spectral and local branches are merged and passed through a nonlinear activation function $\sigma$. Finally, the projection operator $Q$ maps the latent representation to the output field $\mathcal{Y}$.}
    \label{fig:fno}
\end{figure}

The FNO is designed to learn mappings between continuous function spaces, enabling resolution invariant operator learning~\cite{li2020fourier}. In this work, the macro-to-micro parameters mapping is naturally formulated as a nonlinear operator transforming macroscopic plasma observables into spatially resolved EDFs, making FNO well suited for this inverse mapping problem.

The architecture consists of three components, an input lifting layer, a sequence of Fourier operator blocks, and an output projection network as shown in Fig \ref{fig:fno}. The lifting layer maps input fields to a higher-dimensional latent space, which is then processed through multiple Fourier layers to model global spatial dependencies. In each Fourier layer, latent features are transformed to the frequency domain via the Fast Fourier Transform (FFT), where a learnable linear transformation is applied to a truncated set of low-frequency modes. The features are then mapped back to the spatial domain using the inverse FFT. This spectral convolution efficiently captures long-range correlations while maintaining computational efficiency. Each Fourier block also includes a local linear transformation in physical space followed by a nonlinear activation, enabling the model to capture both global structure and localized variations.

We employ a two-dimensional FNO\cite{pandya2026physics} with four Fourier layers and a latent width of 32 channels. Each layer retains 16 Fourier modes along each spatial dimension, focusing on dominant large-scale structures while filtering high-frequency noise. To mitigate boundary artifacts, constant spatial padding of 8 grid cells is applied. Additionally, normalized spatial coordinates are concatenated with the input to preserve positional information during spectral processing. GELU activations are used throughout the network to improve training stability. The final latent representation is mapped to the output space using a pointwise decoder with two fully connected layers of width 64. A sigmoid activation ensures predictions lie within the normalized range $[0,1]$. The model is trained using a loss between predicted and reference EDFs.

\subsection{MeshGraphNet: Graph-Based Relational Learning for Spatially Structured Fields}

\begin{figure}[]
    \centering
    \includegraphics[width=0.7\linewidth]{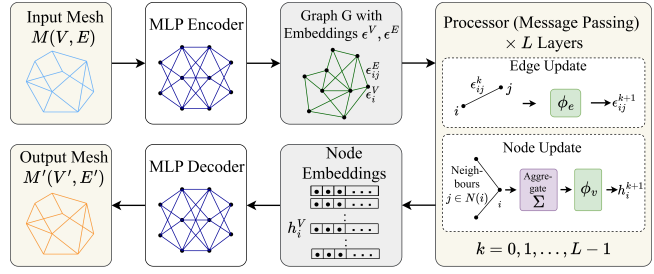}
    \caption{Architecture of the MeshGraphNet model. The input mesh $M(V,E)$, represented by nodes $V$ and edges $E$, is first encoded into latent node and edge embeddings using an MLP encoder. The latent graph is then processed through $L$ message-passing layers, where edge features are updated using the edge-update function $\phi_e$ and node features are updated using the node-update function $\phi_v$ after aggregating information from neighboring nodes. The resulting latent node embeddings are decoded through an MLP decoder to generate predictions on the output mesh $M'(V',E')$.}
    \label{fig:meshgnn}
\end{figure}

MeshGraphNet is a graph neural network architecture designed to learn spatially distributed physical dynamics on mesh-based discretizations through message passing between neighboring nodes~\cite{pfaff2020learning, khemani2024review}. MeshGraphNet represents the computational domain as a graph, where spatial locations are nodes and their physical relationships are encoded as edges. This representation is particularly suitable for mapping from macro to micro plasma parameters, where quantities exhibit spatial gradients and localized variations across the computational domain.

The architecture follows an encode-process-decode framework as shown in Fig. \ref{fig:meshgnn}. In the encoding stage, macroscopic plasma quantities are mapped to node features, while geometric relationships between neighboring locations are represented as edge features. Separate multilayer perceptrons (MLPs) embed node and edge attributes into a higher-dimensional latent space, enabling joint modeling of local states and neighborhood interactions. The processor stage consists of multiple message passing blocks that iteratively update node representations by aggregating information from neighboring nodes. In each step, edge features are first updated based on connected node states, followed by aggregation of neighboring messages to update node embeddings. This iterative process expands the effective receptive field, allowing the model to capture both local and longer-range spatial dependencies.

In this work, the MeshGraphNet operates on graph representations of the two-dimensional grid of plasma parameters, where each grid point is treated as a node connected to its neighbors. The model employs six message passing blocks with a latent feature width of 64 across the node encoder, edge encoder, processor, and decoder modules. Both node and edge encoders use two-layer MLPs with ReLU activation. Following message passing, a node-wise decoder (two-layer MLP) maps latent node representations to the required output channels. A sum-based aggregation scheme ensures permutation-invariant message fusion. A sigmoid activation is applied at the output layer to enforce predictions within the normalized range $[0,1]$. The model is trained using a loss between predicted and reference EDF histograms. By learning relational interactions from graph connectivity, MeshGraphNet enables information to propagate across the domain. 


\subsection{Training Strategy and Experimental Setup}

The training and evaluation framework is designed to assess the ability of the models to generalize across different physical regimes rather than memorizing case-specific patterns. As described in Section~\ref{data generation}, the dataset is generated from seven distinct PIC simulation cases, each corresponding to different plasma conditions. Each case consists of 2001 input-output samples. To evaluate cross-regime generalization, cases 1-5 from Table. \ref{tab:testcases} are used for training and validation, resulting in total of 10005 samples. The combined dataset from the five training cases is shuffled and split into training and validation subsets using a 70:30 ratio. Cases 6 and 7 from Table \ref{tab:testcases} were excluded from the training process and used exclusively for testing. The random 70:30 split is used only for model selection, whereas the two held-out simulation cases provide the independent assessment of cross-condition generalization. This strategy ensures that the trained models are evaluated under physically distinct operating conditions that were not encountered during training, thereby assessing the ability of the learned mapping to generalize to completely unseen regimes.

All three models, U-Net, FNO and MeshGraphNet, are trained under a consistent optimization setup to enable a fair comparison. Each model is trained for 100 epochs using the Adam optimizer with an initial learning rate of $1 \times 10^{-3}$. A learning rate scheduler reduces the learning rate by a factor of 2 if the validation loss does not improve for 10 consecutive epochs. The training objective is defined using the Huber loss function\cite{huber1964robust}, which balances mean squared error and mean absolute error, making it suitable for capturing both large deviations and small fluctuations in the predicted energy distributions. To ensure consistency with the adopted min-max normalization scheme, all models employ a sigmoid activation function at the output layer, constraining predictions within the normalized range $[0,1]$ and preventing unbounded outputs.

\section{Model Evaluation and Results}

\begin{table}[t]
\centering
\caption{Model complexity and computational performance\\
*Inference time is measured as the average time per sample on Test Data (case 6 and case 7)}
{\begin{tabular}{|c|c|c|c|c|}
\hline
\textbf{Model} & \textbf{Parameters} & \textbf{Best Epoch} & \textbf{Inference Time / Sample (ms)$^{*}$} \\
\hline
U-Net & 31,043,716 & 74 & 5.349176\\
\hline
FNO & 4,205,652 & 100 & 0.309875  \\
\hline
MeshGraphNet & 251,076 & 91 & 25.029857 \\
\hline
\end{tabular}
}
\label{tab:model_comparison}
\end{table}

This section presents a comprehensive evaluation of the proposed neural architectures in terms of predictive accuracy, computational efficiency, and physical consistency. The reconstruction is assessed at two complementary levels, first, by comparing the spatially resolved energy distributions with the PIC reference distributions, and second, by evaluating the corresponding spatially integrated EEDF/IEDF. Training was performed on a single NVIDIA RTX 6000 Ada GPU (CUDA 12.4) without imposing any model-specific computational constraints. For each architecture, the checkpoint corresponding to the lowest validation loss was selected for final evaluation. In addition to predictive accuracy, we compare the computational characteristics of the models, including the number of trainable parameters, the epoch corresponding to the best-performing checkpoint, and the average inference time per sample, as summarized in Table \ref{tab:model_comparison}. These metrics provide insight into the trade-off between predictive performance and computational efficiency across the three neural architectures.

\subsection{Evaluation Metrics}

\begin{table}
\centering
\caption{Performance of the proposed models on the validation and testing datasets. The reported testing metrics are computed over the combined electron and ion samples from cases 6 and 7.}
\label{tab:model_results}

\begin{tabular}{|c|c|c|c|c|}
\hline
\textbf{Model} & \textbf{Dataset} & \textbf{MSE}
& \textbf{SSIM} & \textbf{PSNR} \\
\hline

\multirow{2}{*}{U-Net}
& Validation & $1.134 \times 10^{-4}$ & 0.9636 & 39.4545 \\
& Testing    & $1.421 \times 10^{-4}$ & 0.9538 & 38.4751 \\
\hline

\multirow{2}{*}{FNO}
& Validation & $0.332 \times 10^{-4}$ & 0.9811 & 44.7927 \\
& Testing    & $0.880 \times 10^{-4}$ & 0.9738 & 40.5544 \\
\hline

\multirow{2}{*}{MeshGraphNet}
& Validation & $0.370 \times 10^{-4}$ & 0.9805 & 44.3175 \\
& Testing    & $1.284 \times 10^{-4}$ & 0.9708 & 38.9144 \\
\hline

\end{tabular}
\end{table}

Model performance is evaluated using a combination of pixel-wise error metrics and perceptual similarity measures to assess both numerical accuracy and structural fidelity of the reconstructed energy distribution histograms. Specifically, model performance is assessed using mean squared error (MSE), peak signal-to-noise ratio (PSNR), and structural similarity index (SSIM)\cite{wang2004image}. MSE quantifies the average squared deviation between the predicted and reference distributions, thereby emphasizing larger reconstruction errors. PSNR, which is derived from the MSE, provides a logarithmic measure of reconstruction quality, with higher values indicating better agreement between prediction and reference. SSIM complements these metrics by measuring structural similarity through comparisons of luminance, contrast, and local spatial structure, making it particularly suitable for assessing the preservation of spatially meaningful features. Together, these metrics provide complementary assessments of reconstruction quality by evaluating pixel-wise accuracy as well as structural agreement between the predicted and reference distributions.

Table~\ref{tab:model_results} summarizes the validation and testing performance of the three architectures, where the reported metrics are computed over the combined electron and ion datasets from cases 6 and 7. The results exhibit consistent trends between the validation and testing datasets, indicating stable generalization across different plasma conditions. Among the three architectures, FNO achieves the lowest MSE together with the highest PSNR and SSIM, indicating superior reconstruction accuracy and structural fidelity. MeshGraphNet exhibits performance comparable to FNO on the validation set, although its performance degrades slightly on the testing dataset, suggesting reduced generalization under previously unseen simulation conditions. U-Net, while competitive, exhibits higher error values.These results suggest that architectures relying primarily on local convolutional operations may be less effective at learning the highly nonlinear macro-to-micro mapping than architectures capable of modeling global interactions. The quantitative evaluation identifies FNO as the most accurate architecture while also exhibiting the lowest inference time and substantially fewer trainable parameters than U-Net (Table~\ref{tab:model_comparison}), providing an advantageous balance between reconstruction quality and computational efficiency.

\subsection{Qualitative Analysis of Reconstructed EDFs}

Figures~\ref{fig:qualitative_results_electron} and~\ref{fig:qualitative_results_ion} present qualitative comparisons of the reconstructed EEDFs and IEDFs respectively, for an unseen test sample from case 6. Each figure compares the predictions of U-Net, FNO, and MeshGraphNet over the full computational domain as well as the left and bottom sheath regions (as defined in section \ref{data generation}) to the ground truth. The one-dimensional normalized EDFs are obtained by integrating the predicted and reference two-dimensional spatial-energy histograms according to Eq.~\ref{eq:hist_to_dis}, thereby enabling a direct comparison of the reconstructed kinetic distributions. Across all three architectures, the reconstructed EDFs accurately reproduce the dominant characteristics of the reference distributions, including the peak location, distribution width, and overall profile. Agreement is particularly strong within the full domain, where the predicted distributions are nearly indistinguishable from the ground truth. In the sheath regions, where the distributions exhibit stronger gradients and more pronounced non-equilibrium characteristics, slightly larger deviations become visible, particularly in the high-energy tails in IEDF. Since relatively few particles occupy these energy levels, small prediction errors become more apparent after normalization, making the high-energy tails particularly challenging to reconstruct accurately. FNO consistently provides the closest agreement with the reference distributions, while MeshGraphNet achieves comparable performance across most regions. U-Net also captures the overall distribution shape but exhibits comparatively larger deviations in the peak and high-energy tail regions. These qualitative observations are fully consistent with the quantitative metrics reported in Table \ref{tab:model_results} and further demonstrate the ability of FNO and MeshGraphNet to reconstruct both global distribution characteristics and localized kinetic structures. The visual comparisons demonstrate that the proposed models successfully reconstruct physically meaningful energy distributions, with FNO exhibiting the highest visual fidelity across both bulk plasma and sheath regions.
    

\subsection{Physics-Based Validation of the Reconstructed EDFs}

While the preceding metrics evaluate reconstruction quality in the image domain, they do not explicitly quantify agreement between the underlying energy distributions. Therefore, a complementary physics-based evaluation is performed. For an isotropic plasma, the EDF $F(\varepsilon)$ provides the essential information required to recover several important macroscopic quantities to characterize the plasma. Accordingly, in addition to conventional error metrics and visual inspection, we use a validation strategy in which the predicted EDF is used to reconstruct macroscopic plasma quantities. This moment-based validation is particularly suitable for the present problem because small pointwise differences in the distribution do not necessarily translate into significant errors in physically relevant averaged quantities.

The particle density is obtained from the zeroth moment of the EDF,

\begin{equation}
    n_{e,i} = \int_{0}^{\infty} F(\varepsilon)\, d\varepsilon
    \label{density}
\end{equation}

while the mean particle energy is computed from the first moment,

\begin{equation}
    T_{e,i} = \frac{2}{3} n_{e,i} ^{-1}  \int_{0}^{\infty} \varepsilon F(\varepsilon)\, d\varepsilon
    \label{energy}
\end{equation}

\begin{figure}
\centering
\includegraphics[width=0.78\linewidth]{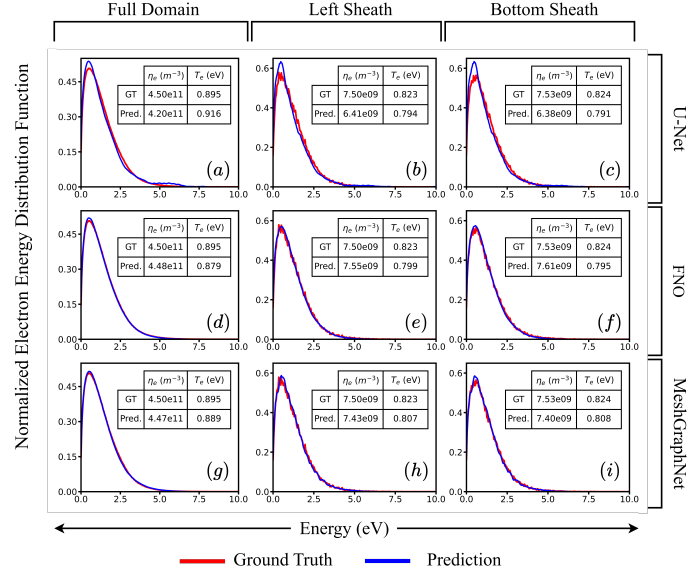}
    \caption{Comparison of the predicted and ground-truth (GT) EEDFs for an unseen sample from case 6 at time $t=1.3425$ \unit{\us}. The $3\times3$ panel layout is organized such that the rows correspond to the U-Net, FNO, and MeshGraphNet models, while the columns represent the full computational domain, left sheath, and bottom sheath regions, respectively. The sheath distributions are obtained from a spatial subset of width $10\Delta x$ extracted from the $256\times256$ computational domain. In each subplot, the ground-truth EEDF is shown in red and the corresponding model prediction in blue. The inset table reports the electron density ($n_e$) and electron temperature ($T_e$) calculated from the ground-truth and predicted (Pred.) distributions, illustrating the ability of the reconstructed EEDFs to recover the corresponding electron density and electron temperature. Electron energy distributions are shown over the range 0-10 eV.}
    \label{fig:qualitative_results_electron}
\end{figure}

\begin{figure}
\centering
\includegraphics[width=0.78\linewidth]{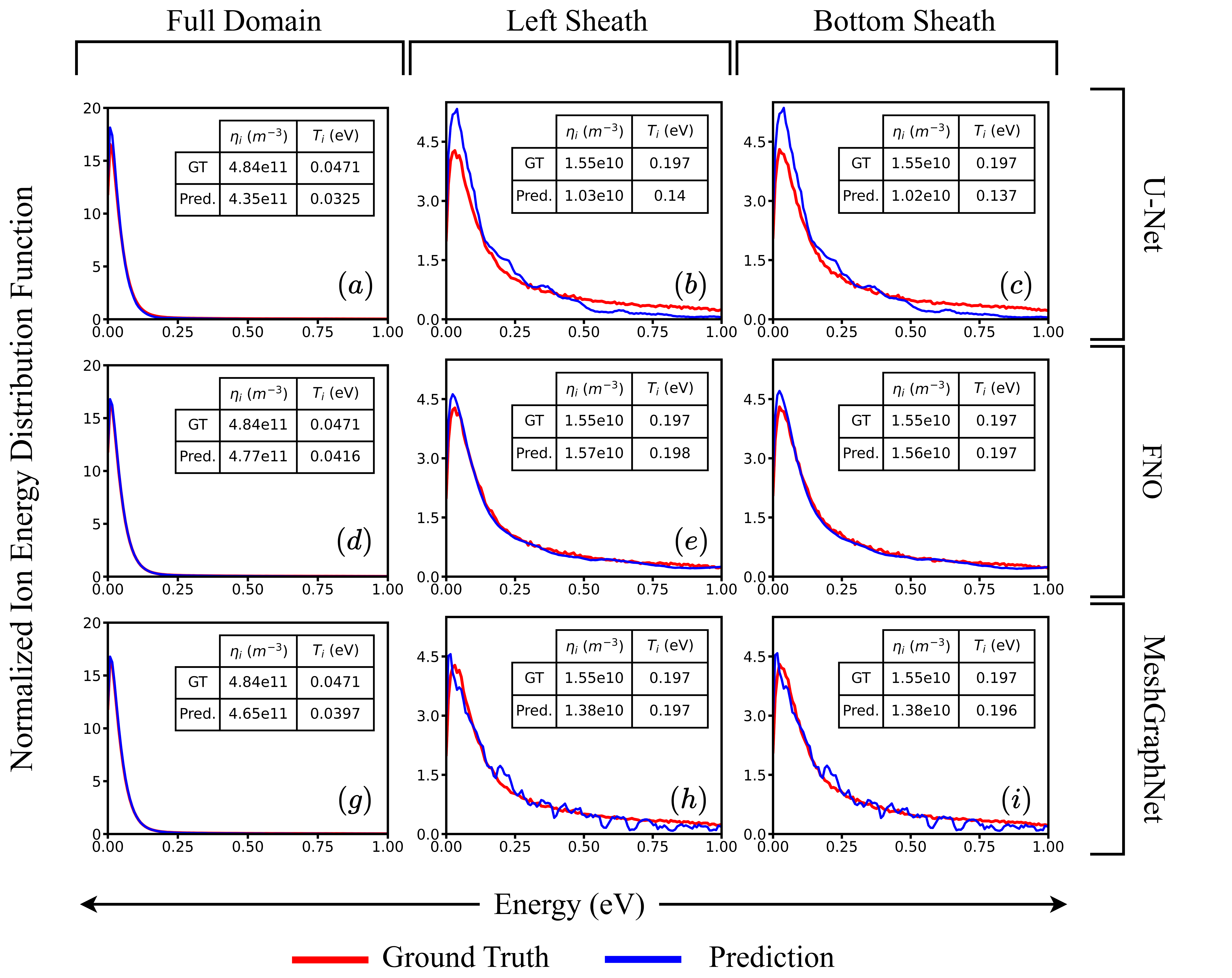}
    \caption{Comparison of the predicted and ground-truth ion energy distribution functions (IEDFs) for the unseen sample from case 6 at time $t=1.3425$ \unit{\us}. The $3\times3$ panel layout is organized such that the rows correspond to the U-Net, FNO, and MeshGraphNet models, while the columns represent the full computational domain, left sheath, and bottom sheath regions, respectively. The sheath distributions are obtained from a spatial subset of width $10\Delta x$ extracted from the $256\times256$ computational domain. In each subplot, the ground-truth IEDF is shown in red and the corresponding model prediction in blue. The inset table reports the ion density ($n_i$) and ion temperature ($T_i$) calculated from the ground-truth (GT) and predicted (Pred.) distributions, illustrating the ability of the reconstructed IEDFs to recover the corresponding electron density and electron temperature. Ion energy distributions are plotted over the range 0-1 eV (zoomed from the full 0-2 eV range) to highlight differences in the low-energy regime.}
\label{fig:qualitative_results_ion}
\end{figure}

The corresponding density and effective temperature calculated from the reconstructed EEDFs and IEDFs are then compared with the corresponding averaged quantities extracted directly from the simulation. This moment-based comparison provides a physically interpretable assessment of the prediction quality and verifies whether the predicted distribution retains the key macroscopic properties of the plasma. Fig. \ref{fig:qualitative_results_electron} and \ref{fig:qualitative_results_ion} suggests that across all three architectures, the reconstructed distributions recover the corresponding macroscopic quantities with good agreement to the ground truth data from PIC simulations for the full simulation domain for both EEDF and IEDF. In particular, the FNO consistently exhibits the smallest deviations in both density and effective temperature, indicating that the learned distributions preserve not only the overall EDF shape but also its physically relevant moments. All models perform optimally for EEDF prediction in the sheath regions; however, for IEDF, where the distributions exhibit stronger non-equilibrium behavior, the predicted macroscopic quantities from the U-Net show small deviations, suggesting that further scope for improvement in model optimization may exist.

\subsection{Validation of the Reconstructed EDFs using electron neutral collision rate coefficients}
The rate coefficients required for characterizing LTP kinetics can be obtained from the EEDF using the following equation.
\begin{equation}
k_{k} = \gamma \int_{0}^{\infty} \varepsilon\, \sigma_{k}\, F(\varepsilon)\, d\varepsilon
\label{eq:ratecoffi}
\end{equation}
Here, $k_{k}$ denotes the rate coefficient for each individual collision process $k$. The term $\gamma$ is a constant equal to $\sqrt{2e/m_{e}}$, where $m_{e}$ is the electron mass and e is the electron charge. The variable $\varepsilon$ represents the electron energy, $\sigma_{k}$ is the energy dependent collision cross section for process $k$, and $F(\varepsilon)$ is the normalized electron EEDF satisfying $\int_{0}^{\infty} F(\varepsilon)\, d\varepsilon = 1$. Details can be found in the work of Hagelaar and Pitchford \cite{hagelaar2005solving}. Using Eq. \ref{eq:ratecoffi}, the momentum transfer rate coefficient, $k_{k=\mathrm{elastic}}$, was evaluated from both the model-predicted and ground-truth EEDFs. For test case 6, at $t=1.3425~\mu\mathrm{s}$, the rate coefficients from FNO-predicted EEDF and ground-truth EEDF were found to be $1.32\times10^{-13}$ and $1.31\times10^{-13}~\mathrm{m^{3},s^{-1}}$, respectively, corresponding to a relative error of only $0.16\%$. When evaluated over the complete simulation duration, the mean percentage errors in the recovered momentum transfer rate coefficient were $0.85\%$, $0.34\%$, and $0.20\%$ for U-Net, FNO, and MeshGraphNet, respectively, for case 6. For test case 7, the corresponding mean errors were $0.94\%$, $0.67\%$, and $0.35\%$ for U-Net, FNO, and MeshGraphNet, respectively. This analysis shows that all three models effectively preserve collision kinetic information in the EEDF with high fidelity, especially in the high-energy tail of the distribution, with MeshGraphNet exhibiting the lowest rate-coefficient error across both test cases.

Additionally, the Jensen-Shannon Divergence (JSD) \cite{lin2002divergence} is used to evaluate the similarity between the two distributions (ground truth and prediction) . Complementing the moment-based validation, the JSD quantifies the similarity between the reconstructed and reference distributions for both EEDF and IEDF in a probabilistic sense. Consistent with the pixel-wise metrics, FNO achieves the lowest JSD values for both electron and ion distributions, indicating the closest agreement with the reference kinetic distributions  (Table \ref{tab:noise_comparison}).

\subsection{Sensitivity to Perturbed Input Fields}

As a controlled stress test and to assess the robustness of the proposed models under imperfect measurement conditions, an additional evaluation is performed using noisy input data. In practical plasma diagnostic systems, measured macroscopic quantities are often associated with experimental uncertainties and sensor noise. Therefore, assessing the sensitivity of the learned macro-to-micro mapping to perturbations in the input plasma observables is important for determining the practical applicability of the proposed framework. For this study, the two unseen test cases, namely cases 6 and 7, are selected and corrupted with additive white Gaussian noise, which provides a simple first-order approximation of random measurement errors. Noise corresponding to 10\% of the signal magnitude is independently added to each input channel, while the target distributions remain unchanged. The trained models are then evaluated directly on these perturbed inputs, without any additional retraining or fine-tuning. 

Table~\ref{tab:noise_comparison} summarizes the performance of U-Net, FNO, and MeshGraphNet on both clean and noisy datasets. All three models exhibit only minor performance variations under noisy conditions, indicating that the learned macro-to-micro mappings are not highly sensitive to moderate perturbations in the input fields. Among the three architectures, FNO demonstrates the strongest robustness to input noise. For both cases 6 and 7, only marginal changes are observed across the MSE, PSNR, SSIM, and JSD metrics for both the electron and ion energy distribution functions after introducing noise. This behavior is consistent with the global spectral representation employed by FNO, which appears to reduce sensitivity to moderate local perturbations in the input fields. MeshGraphNet also maintains relatively stable performance, although it exhibits slightly greater degradation than FNO. The observed robustness is consistent with its message-passing framework, which aggregates information from neighboring nodes and may therefore reduce the influence of localized perturbations. Interestingly, U-Net exhibits only minor performance variations and even slight improvements in several metrics following the introduction of moderate input noise. Such behavior may arise because the imposed perturbations are relatively small and convolutional operations tend to suppress localized fluctuations. Overall, these results demonstrate that the proposed framework maintains high predictive accuracy under moderate input perturbations, indicating that the learned macro-to-micro mapping is robust to moderate input uncertainty.

\begin{table*}[]
\centering
\footnotesize
\setlength{\tabcolsep}{3.5pt}

\begin{tabular}{llcccccccc}
\toprule
\multirow{2}{*}{\textbf{Model}} &
\multirow{2}{*}{\textbf{Case}} &
\multicolumn{4}{c}{\textbf{Electron}} &
\multicolumn{4}{c}{\textbf{Ion}} \\

\cmidrule(lr){3-6}
\cmidrule(lr){7-10}

&
&
\makecell{\textbf{MSE}\\($\times10^{-4}$)} &
\textbf{PSNR} &
\textbf{SSIM} &
\makecell{\textbf{JSD}\\($\times10^{-3}$)} &
\makecell{\textbf{MSE}\\($\times10^{-4}$)} &
\textbf{PSNR} &
\textbf{SSIM} &
\makecell{\textbf{JSD}\\($\times10^{-3}$)} \\

\midrule

\multirow{4}{*}{U-Net}
& 6      & 2.2743 & 36.6204 & 0.9118 & 6.2579 & 0.6004 & 42.5335 & 0.9850 & 5.2439 \\
& $6^{\dagger}$ & 2.0390 & 37.1224 & 0.9132 & 5.8342 & 0.5880 & 42.5245 & 0.9858 & 4.9566 \\
& 7      & 1.9895 & 37.1276 & 0.9275 & 2.7291 & 0.8184 & 41.2157 & 0.9911 & 4.4738 \\
& $7^{\dagger}$ & 1.8878 & 37.4030 & 0.9256 & 2.8140 & 0.7583 & 41.5280 & 0.9920 & 4.2864 \\

\midrule

\multirow{4}{*}{FNO}
& 6      & 0.8457 & 40.9098 & 0.9547 & 0.7577 & 0.1739 & 47.9712 & 0.9962 & 1.3662 \\
& $6^{\dagger}$ & 0.8447 & 40.9010 & 0.9526 & 0.9838 & 0.1921 & 47.5255 & 0.9958 & 1.4016 \\
& 7      & 1.0944 & 39.9167 & 0.9512 & 0.6142 & 1.4065 & 39.4636 & 0.9932 & 2.2071 \\
& $7^{\dagger}$ & 1.0990 & 39.9097 & 0.9512 & 0.6217 & 1.3076 & 39.7152 & 0.9931 & 2.2818 \\

\midrule

\multirow{4}{*}{MeshGraphNet}
& 6      & 0.7943 & 41.0942 & 0.9508 & 0.8923 & 0.9973 & 40.8826 & 0.9913 & 2.6454 \\
& $6^{\dagger}$ & 0.9328 & 40.4081 & 0.9443 & 1.4660 & 1.0424 & 40.6364 & 0.9904 & 2.8628 \\
& 7      & 1.4639 & 38.9266 & 0.9523 & 0.1459 & 1.8803 & 37.7543 & 0.9890 & 4.2878 \\
& $7^{\dagger}$ & 1.5116 & 38.7235 & 0.9509 & 0.1722 & 2.0014 & 37.5073 & 0.9891 & 4.2928 \\

\bottomrule
\end{tabular}

\caption{Performance comparison of U-Net, FNO, and MeshGraphNet on Cases~6 and~7 and their noisy counterparts. Cases marked with $\dagger$ denote inputs corrupted with 10\% additive white Gaussian noise (AWGN), while the target distributions remain unchanged. The reported metrics include Mean Squared Error (MSE), Peak Signal-to-Noise Ratio (PSNR), Structural Similarity Index (SSIM), and Jensen--Shannon Divergence (JSD), evaluated separately for the electron and ion energy distribution functions (the closer the value is to 0, the better the match).}

\label{tab:noise_comparison}

\end{table*}

\subsection{Scope, Generalizability, and Future Directions}
The present work establishes the feasibility of the proposed macro-to-micro reconstruction within the range of plasma conditions represented in the PIC simulation generated datasets. The learned mapping is therefore conditioned on the physical models, collision cross sections, boundary conditions, and parameter space used to generate the training data, and predominantly Maxwellian like energy distributions. Its transferability beyond these assumptions therefore remains to be established. 
Additionally, the present EDF representation assumes an isotropic velocity distribution and consequently does not capture directional information in velocity space. 
Future work will extend the representation to the full velocity space distribution to account for anisotropic kinetic states. Finally, the results demonstrate empirical predictability of the kinetic distribution within the sampled simulation space, but do not establish the mathematical uniqueness of the inverse mapping. Nor do they establish superiority over analytical or physics based mapping approaches, which may provide greater physical interpretability. 
Future studies incorporating experimentally informed uncertainty models and experimental data will also be required to assess practical transferability.

\section{Conclusion}
In this work, we investigated the inverse problem of reconstructing spatially resolved electron and ion energy distribution 
functions (EEDFs and IEDFs) from readily available macroscopic plasma quantities. 
Using paired datasets generated from high-fidelity electrostatic 2D-3V PIC-MCC simulations spanning seven physically distinct simulation conditions, we provide evidence that, within the range of plasma conditions represented in the present dataset, macroscopic plasma observables retain sufficient information to enable an approximate reconstruction of the underlying kinetic state and this nonlinear mappings can be successfully learned using data-driven models. A carefully designed, physics-diverse dataset has been generated spanning multiple LTP discharge conditions, 
providing a representative benchmark for investigating the proposed inverse mapping.
 The reconstructed distributions accurately reproduce both bulk-plasma and sheath characteristics, while physics-based validation 
 confirms that the recovered EDFs preserve the corresponding macroscopic plasma properties, demonstrating that the learned mappings 
 remain physically consistent beyond conventional image-based similarity metrics.

A comparative evaluation of three representative learning paradigms, namely U-Net, Fourier Neural Operator (FNO), and 
MeshGraphNet, showed that all three architectures are capable of approximating the macro-to-micro mapping, with the FNO providing 
the best overall accuracy and generalization across the unseen test cases considered in this study. 
Controlled perturbation tests further demonstrate robustness to moderate levels of measurement noise, suggesting that the proposed framework is not 
overly sensitive to uncertainties typically encountered in LTP diagnostics.

Although the forward relationship between energy distribution functions and macroscopic plasma quantities 
is well established, within the range of plasma conditions represented in the present dataset, this study provides evidence that readily measurable macroscopic plasma quantities retain 
sufficient information to enable an approximate reconstruction of the underlying kinetic state. 
The findings of this study establish a proof of concept and provide a potential foundation for future integration with experimental diagnostics, 
reduced-order kinetic modeling, and other low-temperature plasma applications. More broadly, the proposed framework also illustrates how high-fidelity kinetic simulations can be combined with data-driven models to 
investigate inverse problems that are otherwise difficult to address using conventional analytical or diagnostic approaches, opening 
new opportunities for physics informed surrogate models and a potential building block toward physics-informed digital twins in low temperature plasma science.  Future work will focus on extending the framework to more diverse plasma conditions by expanding the simulation dataset together with directional velocity space information and experimentally informed observations, incorporating temporal dynamics, and exploring hybrid physics-informed learning architectures for assessing the generality and physical interpretability of the proposed framework.


\section*{Acknowledgment}
The authors acknowledge funding support from the ANRF, Government of India (Project Code: CRG/2023/007309). Authors acknowledge Vraj Gandhi for U-NET implementation and Ayushi Sharma for PIC related discussions. The authors also acknowledge the assistance of ChatGPT in improving the spelling, grammar, language, and enhancing the clarity of portions of the manuscript.

\section*{Data Availability Statement}
The data that support the findings of this study are available from the corresponding author upon reasonable request.

\vspace{20pt}
\bibliographystyle{abbrv}
\bibliography{article}

\end{document}